\documentclass[12pt,a4paper]{article}
\usepackage[latin1]{inputenc}
\usepackage{amsmath}
\usepackage{amsfonts}
\usepackage{amssymb}
\usepackage[dvips]{graphicx,color}
\author{C.~D.~Fosco~\footnote{Electronic address: fosco@cab.cnea.gov.ar}
          \\ \\
            {\normalsize\it Centro At{\'o}mico Bariloche - Instituto
            Balseiro,}\\
              {\normalsize\it Comisi{\'o}n Nacional de Energ{\'\i}a
              At{\'o}mica}\\
                {\normalsize\it 8400 Bariloche, Argentina.}}  
\date{}

\title{Chirality and parity in a first-quantized representation}
\begin{document}
\maketitle
\begin{abstract}
\noindent We study the realization of chiral and parity transformations
within a particle-like path-integral representation for Dirac fields,
showing how those transformations can be implemented in a natural way
within the formalism.  We then obtain a representation for the chiral
fermion propagator and determinant within this framework, and also
formulate a way to define the average of the propagator over a random
mass in $d=2n$ dimensions.
\end{abstract}

\section{Introduction}\label{sec:intro} 
Some important Quantum Field Theory objects can be conveniently
represented in terms of particle-like path-integrals, something which
has been known since a long time ago~\cite{schwinger,feynman}. This
may be more appealing, from the physical point of view, than the
standard `second quantized' picture; besides, some approximation
schemes suggest themselves in a clearer fashion. Some successful
applications of these representations appeared, for example, within
the context of the infrared approximation to the evaluation of the
full Dirac propagator in
$QED$~\cite{Karanikas:1995ua,Karanikas:1995zi,Sanchez-Guillen:2002rz}.
They have also been applied to many other interesting problems, since
its use provides a framework which often becomes very convenient for
the introduction of non-standard calculation
techniques~\cite{Kosower:ic,Bern:1987tw,Strassler:1992zr,Schubert}.

For the case of non-zero spin fields, different proposals for the
construction of an integral over first-quantized trajectories have
been advanced.  Since they usually involve different sets of
variables, the task of relating them is far from trivial, unless it is
undertaken at a purely formal level.

In this article, we build on the results presented in~\cite{pintrep},
which dealt with the particular case of a path-integral representation
for Dirac fields, first introduced by Migdal in~\cite{migdal}. We
first consider the realization of symmetries within this framework: in
particular, we show how chiral transformations can be introduced at
the level of the `particle' trajectories that have to be integrated
out in order to derive the propagator.  We then show how chiral
fermion fields can be described within this formulation, and how this
representation may be used to define chiral fermion propagators and
determinants. 

We conclude with the consideration of a random mass, which is a
pathological situation if considered under the light of a
representation based on~\cite{migdal}. It is, however, an interesting
situation, and we show how the problems can be dealt with and a
representation valid for this case be derived.

This article is organized as follows: in section~\ref{sec:paths} we
discuss chiral and parity transformations within the context of
Migdal's representation. In section~\ref{sec:chiral} we present a
derivation of the expressions for the propagator and effective action
of a chiral fermion in an even-dimensional spacetime.  

Section~\ref{sec:random} deals with the definition of the path integral
for the case of a random mass field, and in section~\ref{sec:concl}, we present our conclusions.

\section{Chiral and parity transformations}\label{sec:paths}
Let us briefly review the main properties of the particle
path-integral representation, as presented in~\cite{pintrep}, for the
case of a massive Dirac fermion in $d$ Euclidean dimensions.
To be specific, we consider a theory where the fermionic action,
$S_f$, is defined by:
\begin{equation}\label{eq:defsf}
S_f({\bar\psi},\psi;A) \;=\; \int d^dx \, {\bar\psi} {\mathcal D}
\psi
\end{equation}
where the Dirac operator ${\mathcal D}$, (defined including the mass
term) is given by \mbox{${\mathcal D}\;=\; \not \!\! D + m$}, with
\mbox{$\not \!\! D = \gamma_\mu D_\mu$} and \mbox{$D_\mu = \partial_\mu + i e A_\mu$}.
$A_\mu$ is an Abelian gauge field. For the case of a non-Abelian group
${\mathcal G}$, we replace $i e A_\mu \to g {\mathcal A}_\mu$ in the
covariant derivative.  Here $g$ is the coupling constant for the
non-Abelian gauge theory, and ${\mathcal A}_\mu$ belongs to the Lie
algebra of ${\mathcal G}$ (${\mathcal A}_\mu$ is anti-Hermitian).

The $\gamma$-matrices are chosen to be Hermitian, and to verify the
anticommutation relations \mbox{$\{ \gamma_\mu , \gamma_\nu \} \;=\; 2 \; \delta_{\mu\nu}$}.

When the number $d$ of spacetime dimensions is even ($d = 2 n$), it is
convenient to introduce a $\gamma_s$ matrix, the generalization of $\gamma_5$
to $n \neq 2$, such that $\{\gamma_s , \gamma_\mu \}= 0$, $\forall \mu$. Following the
conventions of~\cite{zinn}, we assume that $\gamma_s$ is Hermitian, and
that $\gamma_s^2 = 1$.

The Dirac propagator $G(x,y;m)$ is the kernel of the inverse of the
operator defining the quadratic form in $S_f$, equation (\ref{eq:defsf}), namely:
\begin{equation}\label{eq:defg}
G(x,y;m)\;=\; \langle \psi(x) {\bar\psi}(y)\rangle \;=\; 
\langle x |{\mathcal D}^{-1} | y \rangle \;,
\end{equation}
where we have adopted Schwinger's convention: $\langle x|{\mathcal T}|y\rangle$
for ${\mathcal T}(x,y)$, the kernel of any operator ${\mathcal T}$ in
coordinate space.  The dependence of $G$ on the mass has been
explicitly written, and we have omitted the spinorial indices,
although it should be evident from the context that $G$ is a matrix
($2\times2$ for $d=2,3$ and $4\times4$ when $d=4$).

When the mass $m$ is constant and positive, a path-integral
representation associated to $G(x,y;m)$, can be introduced by setting:
\begin{equation}\label{eq:defk}
G(x,y;m) \;=\; \int_0^\infty dT \; {\mathcal K}(x,y;m,T)
\end{equation}
where
\begin{equation}\label{eq:kpint}
{\mathcal K}(x,y;m,T)\;=\; \langle x | \exp[ - T ( \not \!\! D +
m) ] \; |y\rangle
\end{equation}
has the path-integral representation:
\begin{equation}\label{eq:ppi}
{\mathcal K}(x,y;m,T)\;=\; \int_{x(0) = y}^{x(T) = x} {\mathcal
D}p {\mathcal D}x \;e^{\int d\tau ( i p \cdot {\dot x} - m)} \Phi(T) \;
e^{- i e \int_0^T d\tau {\dot x}\cdot A} \;,
\end{equation}
with
\begin{equation}\label{eq:defphit}
\Phi(T)\;=\; {\mathcal P}[ e^{- i \int_0^T d\tau {\not p}(\tau)}] \;.
\end{equation}

There is an analogous path-integral like expression for
$\Gamma(A,m)$, the $1$-loop contribution to the effective action due to the
fermions:
\begin{equation}\label{eq:defga}
\Gamma(A,m)\;=\; - \ln \det {\mathcal D} \;,
\end{equation}
where we have omitted an infinite additive constant (that corresponds
to the determinant of the free Dirac operator). The path integral
enters into the game through the representation:
\begin{equation}
\Gamma(A,m) \;=\; \int_0^\infty \frac{dT}{T} \; \int d^d x \; 
{\rm tr}\, {\mathcal K}(x,x;m,T)\;,
\end{equation}
with ${\mathcal K}$ defined as in (\ref{eq:kpint}).

On the other hand, if the mass $m$ is {\em negative\/}, it is still
possible to represent $G(x,y;m)$ in a similar way:
\begin{equation}\label{eq:negativem}
G(x,y;m) \;=\; - \; \int_{-\infty}^0 dT \, 
{\mathcal K}(x,y;m,T) \;,
\end{equation}
while for $\Gamma$ the corresponding formula reads:
\begin{equation}\label{eq:gax}
\Gamma(A,m) \;=\; \int_{-\infty}^0  \frac{dT}{T} \; \int d^d x \; 
{\rm tr}\, {\mathcal K}(x,x;m,T)\;.
\end{equation}

The effect of global chiral transformations on Dirac operators is of
course well known, and can be obtained from the fermionic action
(\ref{eq:defsf}), by performing the corresponding transformations on
the fermionic fields. The chiral properties of ${\mathcal D}$ are
reflected in the effect of a $\gamma_s$ transformation, for example, by
considering:
\begin{equation}
{\mathcal D}' \;=\; \gamma_s {\mathcal D} \gamma_s \;,
\end{equation}
which corresponds to the discrete transformation of the fermionic
fields:
\begin{equation}\label{eq:defctrans}
\psi' (x) \;=\;  \gamma_s \, \psi(x)\;\;,\;\;\; {\bar \psi}'(x) \;=\;
{\bar\psi}(x) \gamma_s \;.
\end{equation}
For the propagator, we have
\begin{equation}\label{eq:g5prop}
\gamma_s [G(x,y;m)]^{-1} \gamma_s\;=\; - [G(x,y;-m)]^{-1} \;.
\end{equation}

Let us see how these transformations are realized at the level of the
path integral representation for $G$, assuming $m>0$. Clearly, the
only effect of the transformations (\ref{eq:defctrans}) is to change
the sign of the exponent of the path-ordered exponential:
$$
\gamma_s G(x,y;m) \gamma_s \;=\; \int_0^\infty dT \, \int_{x(0) = y}^{x(T) = x}
{\mathcal D}p {\mathcal D}x \; e^{\int_0^T d\tau [ i p\cdot{\dot x} - m ]}
$$
\begin{equation}\label{eq:tpi1}
\times {\mathcal P}[ e^{+ i \int_0^T d\tau {\not p}}] \; e^{- i e \int_0^T
d\tau {\dot x}\cdot A} \;.
\end{equation}
To relate the previous expression to the path integral representation
for $G$, we perform a reparametrization of the variables; $\tau \to -\tau$,
under which
\begin{equation}
{\tilde p} (\tau) \;=\; p (-\tau) \;\;,\;\; {\tilde x} (\tau)
\;=\; x (-\tau) \;,
\end{equation}
and
$$
\gamma_s G(x,y;m) \gamma_s \;=\; \int_0^\infty dT \, \int_{{\tilde x}(0) = y}^{{\tilde
    x}(-T) = x} {\mathcal D}{\tilde p} {\mathcal D}{\tilde x} \;\;
e^{\int_0^{-T} d\tau [ i {\tilde p}\cdot{\dot {\tilde x}} + m ]}
$$
\begin{equation}\label{eq:tpi2}
{\mathcal P}[ e^{- i \int_0^{-T} d\tau {\not {\tilde p}}}] \;\; e^{- i e \int_0^{-T}
d\tau {\dot {\tilde x}}(\tau)\cdot A[{\tilde x}]} \;.
\end{equation}
Then we change variables: $T \to -T$ in the integration over $T$, to
obtain:
$$
\gamma_s G(x,y;m) \gamma_s \;=\; - \int_0^{-\infty} dT \, \int_{{\tilde x}(0) =
  y}^{{\tilde x}(T) = x} {\mathcal D}{\tilde p} {\mathcal D}{\tilde x}
\; e^{\int_0^T d\tau [ i {\tilde p}\cdot{\dot {\tilde x}} + m ]}
$$
\begin{equation}\label{eq:tpis}
{\mathcal P}[ e^{- i \int_0^T d\tau {\not {\tilde p}}}] \;\; e^{- i e
\int_0^T
d\tau {\dot {\tilde x}}(\tau)\cdot A[{\tilde x}]} \;,
\end{equation}
or:
\begin{equation}
\gamma_s \; G(x,y;m) \; \gamma_s \;=\; \int_{-\infty}^0 \; dT \;
{\mathcal K}(x,y;-m,T)\;.
\end{equation}
Recalling (\ref{eq:negativem}), and that we had assumed that $m>0$, we
conclude that (\ref{eq:g5prop}) holds true.

On the other hand, parity transformations in an odd-dimensional
Euclidean spacetime can be represented by:
\begin{equation}
\psi(x) \;\to\; i \psi (-x) \;\;\;
{\bar\psi} \;\to \; i {\bar\psi} (-x) \;,
\end{equation}
so that the parity transformed of ${\mathcal D}$ differs only in the
sign of its mass with ${\mathcal D}$.  For the propagator, a similar
calculation to the one for the $\gamma_s$ case leads to a 
relation between $G$ and its parity transformed $G'$:
\begin{equation}
[  G(x,y;m) ]' \;=\; - \int_{-\infty}^0 \; dT \;
{\mathcal K}(x,y;-m,T)\; = \;  G(x,y;-m) \;,
\end{equation}
while for $\Gamma$ we have
\begin{equation}\label{eq:gprima}
[\Gamma(A,m)]' \;=\; \int_{-\infty}^0 \frac{dT}{T} \; \int d^d x \;
{\rm tr}\, {\mathcal K}(x,x;-m,T)\;.
\end{equation}
We can derive similar relations involving $\gamma_s$ transformations in
even numbers of dimensions. For example, due to the
cyclic property of the trace, in $d=2n$ we have
\begin{equation}\label{eq:geven}
\Gamma_{d=2n}(A,m)\;=\; \int_0^\infty  \frac{dT}{T} \; \int d^d x \; {\rm tr}\, \left[ \gamma_s {\mathcal
K}(x,x;m,T) \gamma_s\right] \;, 
\end{equation}
and by a similar argument to the one used in (\ref{eq:tpis}), we see
that the only change appears in the sign of $p$ in the path-ordered
factor. Thus we may write an equivalent but more symmetric expression
\begin{equation}\label{eq:gsym}
\Gamma_{d=2n}(A,m)\;=\; \int_0^{\infty} \frac{dT}{T} \; \int d^d x \; {\rm tr}\, \left[{\mathcal
K}_{sym}(x,x;m,T)\right] \;, 
\end{equation}
with:
\begin{equation}\label{eq:ksym}
{\mathcal K}_{sym}(x,y;m,T)\;=\; \int_{x(0) = y}^{x(T) = x} {\mathcal
D}p {\mathcal D}x \;e^{\int d\tau ( i p \cdot {\dot x} - m)} {\Re [\Phi(T)]} \;
e^{- i e \int_0^T d\tau {\dot x}\cdot A} \;,
\end{equation}
where the real part of $\Phi(T)$ may of course be also represented as:
\begin{equation}
{\Re [\Phi(T)]} \;=\; {\mathcal P} \cos \left[  \int_0^T d\tau {\not \! p}(\tau)\right] \;.
\end{equation}

On the other hand, we can also obtain a representation for the
parity-odd part of the effective action (in an odd number of spacetime
dimensions). This object, $\Gamma_{odd}$, is given explicitly by:
\begin{equation}\label{eq:godd}
\Gamma_{odd}(A,m)\;=\; \int_0^{\infty} \frac{dT}{T} \; \int d^d x \; {\rm tr}\, \left[{\mathcal
K}_{odd}(x,x;m,T)\right] \;, 
\end{equation}
with
\begin{equation}\label{eq:ksym1}
{\mathcal K}_{odd}(x,y;m,T)\;=\; i\, \int_{x(0) = y}^{x(T) = x} {\mathcal
D}p {\mathcal D}x \;e^{\int d\tau ( i p \cdot {\dot x} - m)}  {\Im [\Phi(T)]} \;
e^{- i e \int_0^T d\tau {\dot x}\cdot A} \;.
\end{equation}
$\Gamma_{odd}$ is proportional to the imaginary part of the effective
action, since $\Gamma_{odd} = i \Im (\Gamma)$.

Taking advantage of (\ref{eq:gprima}), for any
value of the sign of $m$, we have:
\begin{eqnarray}
\Gamma_{odd} &=& \frac{1}{2} {\rm sign} (m) \, \left[ \int_0^\infty \frac{dT}{T} \, 
\int d^d x \, {\rm tr}\, {\mathcal K}(x,x;m,T) \right. \nonumber\\
&-& \left. \int_{-\infty}^0 \frac{dT}{T} \,\int d^d x \, {\rm tr}\, {\mathcal
K}(x,x;- m,T)\right] \;,
\end{eqnarray}
or:
\begin{equation}
\Gamma_{odd}\;=\;{\rm sign} (m) \int_{-\infty}^{\infty}dT \; \frac{ e^{-|m T|}}{2 |T|} 
\, \int d^d x \, {\rm tr} \, {\mathcal K}(x,x;0,T) \;. 
\end{equation}
Although this is in principle valid for any constant value of the
mass, the difficulties to generalize the result to dynamical or random
masses should be evident. This problem is discussed in section~\ref{sec:random}.

The generalization of the previous expressions to the non-Abelian case amounts 
to replacing everywhere the Wilson line factor for the corresponding non-Abelian 
object. Namely,
\begin{equation}\label{eq:nona}
\exp[ - i e \int_0^T d\tau {\dot x}_\mu(\tau) A_\mu(x(\tau)) ] \to
{\mathcal P} \exp[ -  g \int_0^T d\tau {\dot x}_\mu(\tau) {\mathcal A}_\mu(x(\tau)) ] \;.
\end{equation}

\section{Chiral fermions}\label{sec:chiral} 
 The matrix $\Phi(T)$, defined in (\ref{eq:defphit}), verifies the
`evolution equation':
\begin{equation}\label{eq:phieq}
i \partial_T \Phi (T) \;=\; \not \! p(T) \Phi(T) \;\;\; T \,\in\, [ 0, \infty) \;.
\end{equation}
To solve it for $\Phi(T)$, we introduce its four `chiral' components, in
the following way
\begin{eqnarray}
\Phi_{LL}(T) &=& P_L\Phi(T)P_L\;,\;\; \Phi_{LR}(T) \;=\; P_L\Phi(T)P_R \nonumber\\
\Phi_{RL}(T) &=& P_R\Phi(T)P_L \;,\;\; \Phi_{RR}(T) \;=\; P_R\Phi(T)P_R \;,
\end{eqnarray}
where $P_{\stackrel{L}{R}} \equiv \frac{1 \pm \gamma_s}{2}$. 

Of course, we have:
\begin{equation}\label{eq:decomp1}
\Phi(T) \;=\; \Phi_{LL}(T) \,+\, \Phi_{LR}(T)\,+\, \Phi_{RL}(T)\,+\, \Phi_{RR}(T) \;,
\end{equation}
and this kind of decomposition holds also true for the propagator and
effective action, since both depend linearly on $\Phi(T)$.

The equation of motion (\ref{eq:phieq}) is then equivalent to a system
of four equations:
\begin{eqnarray}
i \partial_T \Phi_{LL}(T) &=&  P_L\not\!p (T)  \Phi_{RL}(T) \;,\;\;
i \partial_T \Phi_{LR}(T) \;=\; P_L\not\!p (T)   \Phi_{RR}(T) \nonumber\\
i \partial_T \Phi_{RL}(T) &=& P_R\not\!p (T)   \Phi_{LL}(T)\;,\;\;
i \partial_T \Phi_{RR}(T) \;=\; P_R\not\!p (T)  \Phi_{LR}(T) \;,
\end{eqnarray}
which form two decoupled pairs; namely $\Phi_{LL}$ only mixes with
$\Phi_{RL}$, and $\Phi_{RR}$ with $\Phi_{LR}$.  For a {\em Dirac\/} field, the
initial condition is $\Phi(0)=1$; it implies $\Phi_{LL}(0) = P_L$,
$\Phi_{RR}(0) = P_R$, and \mbox{$\Phi_{LR}(0) = \Phi_{RL}(0) = 0$}.

The solution to these equations may be found by transforming each pair
of coupled equations into an equivalent system of (coupled) integral
equations. The fact that there is no coupling between pairs,
allows for the introduction of chiral fermions, when the proper
initial conditions for the $\Phi$'s are introduced.

For example, from the first and third differential equations, plus the
initial conditions $\Phi_{LL}(0) = 1$ and $\Phi_{RL}=0$, we obtain: 
\begin{eqnarray}\label{eq:phill}
\Phi_{LL}(T)&=& P_L \,-\, i \int_0^T d\tau \, P_L \not\!p (\tau)  \Phi_{RL}(\tau)
\nonumber\\
\Phi_{RL}(T)&=& -\, i \int_0^T d\tau \, P_R \not\!p (\tau)  \Phi_{LL}(\tau) \;. 
\end{eqnarray}
This allows us to solve for $\Phi_{LL}(T)$ and $\Phi_{RL}(T)$, while the other
two components, $\Phi_{RR}(T)$ and $\Phi_{LR}(T)$, vanish identically for all
$T$ when the initial conditions $\Phi_{RR}(0) = 0$ and $\Phi_{LR}=0$ are
imposed.  Note that these two conditions could be inconsistent on a
Dirac fermion. Besides, note that we use the term `initial' here for
the $T$ evolution, which is completely different to the
time ($x_0$) evolution.

From (\ref{eq:phill}), we then derive an equation involving only $\Phi_{LL}$
\begin{equation}
\Phi_{LL}(T)\;=\; P_L \,-\, \int_0^T d\tau \int_0^\tau d\tau' P_L \not\!p (\tau) P_R
\not\!p(\tau') \, \Phi_{LL}(\tau') \;,  
\end{equation}
which may be formally solved in terms of a series:
$$ \Phi_{LL}(T)\;=\;  P_L \,-\, \int_0^T d\tau \int_0^\tau d\tau' \Lambda^{(+)}_{\mu\nu} p_\mu(\tau)
p_\nu(\tau') P_L \ldots$$ 
$$+\; (-1)^n  \int_0^T d\tau_1 \int_0^{\tau_1} d\tau_2 \ldots \int_0^{\tau_{2n-2}} d\tau_{2n-1} \int_0^{\tau_{2n-1}}
d\tau_{2n}  $$
\begin{equation}
\Lambda^{(+)}_{\mu_1\mu_2} \, p_{\mu_1}(\tau_1) \, p_{\mu_2}(\tau_2) \;\ldots\; \Lambda^{(+)}_{\mu_{2n-1}\mu_{2n}} \,p_{\mu_{2n-1}}(\tau_{2n-1})
\, p_{\mu_{2n}}(\tau_{2n}) P_L \;+\; \ldots 
\end{equation}
where $\Lambda^{(+)}_{\mu\nu} \equiv P_L \gamma_\mu \gamma_\nu $. In $1+1$ dimensions,
$\Lambda^{(+)}_{\mu\nu} \equiv \delta_{\mu\nu} + i \epsilon_{\mu\nu}$, so that $\Phi_{LL}(T)$ may be found,
in that particular case, by solving equations that involve only scalar
and pseudoscalar functions.

Let us examine the particular case of a free ($e=0$) propagator. The functional
integral over $x_\mu$ in ${\mathcal K}$ yields a $\delta$ function of ${\dot
p}_\mu$. In this situation, ${\Phi}_{LL}(T)$ may be explicitly evaluated:
\begin{equation}
\Phi_{LL}(T) \;=\; \cos (T p ) P_L \;,
\end{equation}
where $p = \sqrt{p_\mu p_\mu}$. For ${\tilde G}_{LL}(p)$, the
corresponding component of the propagator, an application of
(\ref{eq:defk}), (\ref{eq:kpint}) and (\ref{eq:ppi}) (in Fourier space) yields:
\begin{equation}\label{eq:gllfree}
{\tilde G}_{LL}(p) \;=\; \int_0^\infty dT \, e^{- m T} \, \cos( p T) P_L \;=\;
\frac{m}{p^2 + m^2} P_L \;, 
\end{equation}
as it should be.

For $\Phi_{RL}(T)$, on the other hand, we obtain: 
\begin{equation}
\Phi_{RL}(T)\;=\; -\, i \,\frac{\not\!p}{p}  \sin( p T) P_L
\end{equation}
which yields:
\begin{equation}\label{eq:grlfree}
{\tilde G}_{RL}(p) \;=\; - i \frac{\not\!p}{p} \, \int_0^\infty dT \, e^{- m T} \, \sin( p T) \;=\;
\frac{- i \not \! p}{p^2 + m^2} P_L \;, 
\end{equation}
which becomes the chiral fermion propagator in the $m\to0$ limit.

The chiral fermion propagator, $G_{RL}$, is then obtained in the general ($e\neq 0$)
case, by considering only the $\Phi_{RL}$ component in the path-integral
for ${\mathcal K}$ and taking the $m \to 0$ limit; namely:
\begin{equation}
G_{RL}(x,y) \;=\; \lim_{m \to 0} \int_0^\infty dT \, {\mathcal K}_{RL}(x,y;m,T)
\end{equation}
where
\begin{equation}
{\mathcal K}_{RL}(x,y;m,T)\;=\; \int_{x(0) = y}^{x(T) = x} {\mathcal
D}p {\mathcal D}x \;e^{\int d\tau ( i p \cdot {\dot x} - m)} \Phi_{RL}(T) \;
e^{- i e \int_0^T d\tau {\dot x}\cdot A} \;.
\end{equation}

It is interesting to check that, when $m \to 0$, $G_{LL}\to 0$, while
$G_{RL}$ tends to the free chiral propagator. Regardless of the fact
that $G_{LL}\to 0$, $\Phi_{LL}$ has a non-zero limit for $m\to0$. This is
related, in our context, to the fact that in order to regulate a
chiral fermion one has to introduce both chiralities. Indeed, if the
UV divergences that arise for small values of $T$ are regulated by
distorting the integral over $T$, this will generally introduce a
non-vanishing $G_{LL}$.

An entirely analogous derivation allows one to define a propagator
for the  opposite chirality. Indeed, one imposes now the
conditions $\Phi_{LL}(0) = \Phi_{RL}(0) = 0$. Then:
\begin{equation}
\Phi_{RR}(T)\;=\; P_R \,-\, \int_0^T d\tau \int_0^\tau d\tau' P_R \not\!p (\tau) P_L
\not\!p(\tau') \, \Phi_{RR}(\tau') \;,  
\end{equation}
and
\begin{equation}
\Phi_{LR}(T)\;=\; -\, i \int_0^T d\tau \, P_L \not\!p (\tau)  \Phi_{RR}(\tau)\;.
\end{equation}

Regarding $\Phi_{RR}(T)$, we have the integral equation:
\begin{equation}\label{eq:eqphir}
\Phi_{RR} (T)\;=\; 1 \,-\, \int_0^T d\tau \int_0^\tau d\tau' p_\mu(\tau) \Lambda^{(-)}_{\mu\nu} p_\nu(\tau')
\Phi_{RR}(\tau') \;,
\end{equation}
with $\Lambda^{(-)}_{\mu\nu} \equiv P_R \gamma_\mu \gamma_\nu$ (in $1+1$ dimensions, $\Lambda^{(-)}_{\mu\nu} \equiv
\delta_{\mu\nu} - i \epsilon_{\mu\nu}$). It of course also admits a formal series solution
like the one for $\Phi_{LL}$, although with the substitution: $\Lambda^{(+)} \to
\Lambda^{(-)}$.

The formal series that solve the equations for $\Phi_{LL}$ and $\Phi_{RR}$,
relevant to the construction of the two  chiral propagators,
can be written in terms of path-ordering operators; indeed:
\begin{eqnarray}
\Phi_{LL}(T) &=& {\mathcal P} \sum_{n=0}^\infty \frac{(-1)^n}{(2 n)!} \left[ \int_0^T d\tau \int_0^T d\tau' p_\mu(\tau)
\Lambda^{(+)}_{\mu\nu} p_\nu(\tau') \right]^n P_L   \nonumber\\
\Phi_{RR}(T) &=& {\mathcal P} \sum_{n=0}^\infty \frac{(-1)^n}{(2 n)!} \left[ \int_0^T d\tau \int_0^T d\tau' p_\mu(\tau)
\Lambda^{(-)}_{\mu\nu} p_\nu(\tau') \right]^n P_R \;. 
\end{eqnarray} 
A square root can be introduced, just in order to have more
compact expressions:
\begin{eqnarray}
\Phi_{\stackrel{LL}{RR}} (T) \;=\; {\mathcal P} \cos \sqrt{ \int_0^T d\tau \int_0^T d\tau' p_\mu(\tau)
\Lambda^{(\pm)}_{\mu\nu} p_\nu(\tau')}\;\; P_{\stackrel{L}{R}}\;.
\end{eqnarray} 
It should be noted that, in the previous expressions, the ${\mathcal
  P}$ operator acts also on the functions that are contracted with a
matrix $\Lambda^{(\pm)}_{\mu\nu}$; for example:
$$
{\mathcal P} \left[ p_\mu(\tau) \Lambda^{(\pm)}_{\mu\nu} p_\nu(\tau')\right] \;=\; \theta(\tau-\tau') p_\mu(\tau)
\Lambda^{(\pm)}_{\mu\nu} p_\nu(\tau') 
$$
\begin{equation}
+\; \theta(\tau'-\tau) p_\mu(\tau') \Lambda^{(\pm)}_{\mu\nu} p_\nu(\tau)\;.
\end{equation}
For a general product, the ${\mathcal P}$ operator acts by ordering
the $p's$ according to their arguments, and connecting them pairwise
with the $\Lambda$ matrix (i.e., connecting each consecutive pair of $p's$).

In $1+1$ dimensions, we can make use of the identities:
\begin{equation}
p_\mu(\tau) \Lambda^{(+)}_{\mu\nu} p_\nu(\tau') \;=\; {\bar p}(\tau) p(\tau')\;\;,\;\;\;
p_\mu(\tau) \Lambda^{(-)}_{\mu\nu} p_\nu(\tau') \;=\; p(\tau) {\bar p}(\tau')
\end{equation}
where $p \equiv p_0 + i p_1$ and ${\bar p} \equiv p_0 - i p_1$, to write the
sum of the series in the following way:
\begin{eqnarray}
\Phi_{LL}(T) &=& {\mathcal P} \cos \sqrt{ \int_0^T d\tau \int_0^T d\tau'  {\bar p}(\tau)
p(\tau')} \; P_L \nonumber\\
\Phi_{RR}(T) &=& {\mathcal P} \cos \sqrt{ \int_0^T d\tau \int_0^T d\tau'  p(\tau) {\bar
p}(\tau')} \; P_R \;.
\end{eqnarray} 

We conclude this section by presenting the expression for effective
actions $\Gamma_{L,R}$ corresponding to the chiral fermion determinant of
the given chirality:
\begin{equation}
\Gamma_{\stackrel{L}{R}} (A) \;=\; \lim_{m \to 0} \int_0^\infty \frac{dT}{T} \; \int d^d x \; 
{\rm tr}\, {\mathcal K}_{\stackrel{LL}{RR}} (x,x;m,T)\;,
\end{equation}
with 
\begin{equation}
{\mathcal K}_{\stackrel{LL}{RR}} (x,y;m,T)\;=\; \int_{x(0) = y}^{x(T) = x} {\mathcal
D}p {\mathcal D}x \;e^{\int d\tau ( i p \cdot {\dot x} - m)}
\Phi_{\stackrel{LL}{RR}} (T) \; e^{- i e \int_0^T d\tau {\dot x}\cdot A} \;.
\end{equation}

The decomposition of $\Phi(T)$ into its components can also be applied to
the derivation of the effect of continuous chiral transformations in
this representation. Indeed, the transformations:
\begin{equation}
\psi(x) \to e^{i \alpha(x) \gamma_s} \psi(x)
\;\;,\;\;
{\bar \psi}(x) \to {\bar\psi}(x) e^{i \alpha(x) \gamma_s}  
\end{equation}
lead to the following transformations at the level of $\Phi(T)$:
\begin{eqnarray}
\Phi_{LL}(T) &\to& e^{i \alpha(x)} \Phi_{LL}(T) e^{i \alpha(y)}\;\;\;\; \Phi_{RL}(T) \;\to\;
e^{-i \alpha(x)} \Phi_{RL}(T) e^{i \alpha(y)} \nonumber\\
\Phi_{RR}(T) &\to& e^{-i \alpha(x)} \Phi_{RR}(T) e^{-i \alpha(y)}\;\;\; \Phi_{LR}(T) \;\to\; e^{i \alpha(x)} \Phi_{LR}(T) e^{-i \alpha(y)}
\end{eqnarray} 
where $x$ and $y$ denote the arguments of ${\mathcal K}(x,y;m,T)$. 

\section{Random mass}\label{sec:random} 
It is interesting to see how a random mass in $d$ dimensions may
also be described within this formalism. By `random mass' we mean that
the mass $m$ in the Dirac operator is now a function $M(x)$, and that one
wants to functionally average over the configurations $M(x)$, with a
weight function. To be concrete, we will consider Gaussian averages $\langle\ldots \rangle_M$
defined by:
\begin{equation}
\langle \ldots \rangle_M \;=\; \int{\mathcal D}M \; \ldots \; e^{- \frac{1}{2g} \int d^dx [ M(x)]^2}
\end{equation}
where $g$ is a positive constant. 
 The main difficult to confront from the point of view of the
 path-integral representation is that a varying mass will take both
 positive and negative values, and the representation we are using
 depends explicitly on the sign of the mass. A na\"\i ve application of
 the previous average formula to the representation for a positive
 mass, say, yields a contribution that blows up when integrating over
 $T$. 

 One way to avoid this problem, is to consider instead the inverse
 of \mbox{${\mathcal H}_D \equiv \gamma_s {\mathcal D}$}. The inverse of
 ${\mathcal H}_D$ is obviously related to the one of ${\mathcal D}$, 
since \mbox{${\mathcal D}^{-1} = {\mathcal H}_D^{-1}  \gamma_s$}.  However,
it is much easier to average over the inverse of ${\mathcal H}_D$.
To see this, we consider the representation:
\begin{equation}
\langle x| {\mathcal H}_D^{-1} |y\rangle \;=\; i \, \int_0^\infty dT \, \langle x | U(T)| y \rangle 
\end{equation}
where $U(T)$ is the evolution operator {\em in real
  time\/} for ${\mathcal H}_D$, which is a Dirac {\em Hamiltonian\/} in
$d$ spatial dimensions. More explicitly
\begin{equation}
{\mathcal H}_D \;=\; \gamma_s ( \not \!\! D + M(x))
\end{equation}
and we assume a $-i \varepsilon$ is added to ${\mathcal H}_D$ to pick up the
proper contribution in the $T$ integral.

Note that $\gamma_s$ plays the role of Dirac's `$\beta$' matrix, while the `$\alpha$'
matrices are given by: $\gamma_s \gamma_\mu = i \alpha_\mu$.   
The kernel ${\mathcal K}$ for $U(T)$  can then be represented
by a path integral:
$$
{\mathcal K}(x,y;M,T)\;=\; \langle x| U(T) |y\rangle 
$$
\begin{equation}
=\;  \int_{x(0) = y}^{x(T) = x} {\mathcal
D}p {\mathcal D}x \;e^{ i \int d\tau  p \cdot {\dot x} } \Phi(T) \;
e^{- i e \int_0^T d\tau {\dot x}_\mu A_\mu} \;,
\end{equation}
where 
\begin{equation}
\Phi(T)\;=\; {\mathcal P} e^{ -i \int_0^T d\tau [ \alpha_\mu p_\mu(\tau) + \gamma_s M(x(\tau))]} \;,
\end{equation}
and the $\mu$ index runs from $0$ to $d-1$. The $\Phi$ factor contains the
mass multiplied by a Hermitian matrix, and is so in a position
analogous to the momentum.  

The average over the mass can now be performed by introducing a local
representation for the factor $\Phi(T)$, like the one considered
in~\cite{pintrep}, obtained by adding Grassmann variables.  For the
interesting  case of $2$ dimensions, the expression is simple, since
\begin{equation}
G(x,y;M)= i \gamma_s
\int_0^\infty dT  \int_{x(0) = y,{\bar\xi}(0)={\bar\xi}}^{x(T) = x,\xi(T)=\xi} {\mathcal D}p {\mathcal
  D}x {\mathcal D}\xi {\mathcal D}{\bar\xi} 
\; \exp[ - {\mathcal S}(p,x,\xi,{\bar\xi};T)]
\end{equation}
where
\begin{eqnarray}
{\mathcal S}(p,x,\xi,{\bar\xi};T)&=& {\bar\xi}(T) \xi(T) + \int_0^Td\tau
\left[- i p\cdot {\dot x} \right. \nonumber\\
&+& {\bar\xi} {\dot\xi} -i M(x(\tau)) ( \xi {\bar\xi} - {\bar\xi} \xi )
- i  (p_0 + i p_1) \xi \nonumber\\
&-& \left. i  (p_0 - i p_1) {\bar\xi}
+ i e {\dot x}\cdot A \right] \;.
\end{eqnarray}
The presence of an imaginary exponent allows us to
average over $M$, obtaining:
\begin{eqnarray}
\langle G(x,y;M)\rangle_M &=& i
\int_0^\infty dT  \int_{x(0) = y,{\bar\xi}(0)={\bar\xi}}^{x(T) = x,\xi(T)=\xi} {\mathcal D}p {\mathcal
  D}x {\mathcal D}\xi {\mathcal D}{\bar\xi} \nonumber\\
&\times & \exp[ - {\mathcal S}_{eff} (p,x,\xi,{\bar\xi};T)]
\end{eqnarray}
where
$$
{\mathcal S}_{av}(p,x,\xi,{\bar\xi};T)\;=\; {\bar\xi}(T) \xi(T) + \int_0^Td\tau
\left[- i p\cdot {\dot x} \right. + {\bar\xi} {\dot\xi} - i  (p_0 + i p_1) \xi 
$$
\begin{equation}
- \left. i  (p_0 - i p_1) {\bar\xi}+ i e {\dot x}\cdot A \right]
- 2 g \int_0^T d\tau \int_0^T d\tau'  ({\bar\xi}\xi) (\tau)  \delta(x(\tau) - x(\tau'))
  ({\bar\xi}\xi) (\tau')\;. 
\end{equation}

\section{Conclusions}\label{sec:concl}
We have derived expressions for the realization of parity and chiral
transformations within a particle-like representation, and for the propagator and determinant of a chiral fermion field. Although we presented the results for the case of an Abelian gauge field, their non-Abelian generalizations proceed simply by making the substitution of equation (\ref{eq:nona}).

We wish to point out that, when one considers a propagator in an external (non-dynamical) field, a particle-like 
representation is more natural than a second quantized functional integral. The reason is that, to represent that kind of propagator, the path integral cannot contain physical {\em field\/} degrees of freedom. It is rather an integral over  particle-like trajectories. Indeed, $G(x,y;m)$ can be obtained from a path integral over Grassmann
fields $\psi$, ${\bar\psi}$ as follows:
\begin{equation}\label{eq:secq}
G(x,y;m) \;=\; [ \det{\mathcal D}]^{-1} \, \int {\mathcal D} \psi {\mathcal
D}{\bar\psi} \; \psi(x)\, \psi(y)\;  e^{ - \int d^dx {\bar\psi} {\mathcal D} \psi } \;,
\end{equation}
where the $\det^{-1}$ factors must be present in order to cancel the
$\det$ which comes from the Grassmann integral. Of course, this is not
a {\em local\/} representation, precisely because of the determinant.
In order to localize it, one is forced to introduce bosonic spinorial
fields $\varphi$, ${\bar\varphi}$, so that:
\begin{equation}\label{eq:secq1}
G(x,y;m) \;=\; \int {\mathcal D} \psi {\mathcal D}{\bar\psi} {\mathcal D}\varphi
{\mathcal D}{\bar \varphi} \; \psi(x)\, {\bar \psi}(y)\;  
e^{ - \int d^dx \left( {\bar\psi} {\mathcal D} \psi  \,+\, {\bar \varphi} {\mathcal
D} \varphi \right)} \;.
\end{equation}
And finally, one can write the even more symmetric expression:
$$
G(x,y;m) \;=\; \int {\mathcal D} \psi {\mathcal D}{\bar\psi} {\mathcal D}\varphi
{\mathcal D}{\bar \varphi} \;\frac{1}{2}[ \psi(x)\, {\bar \psi}(y) + \varphi(x) {\bar\varphi}(y)]
$$
\begin{equation}
e^{ - \int d^dx \left( {\bar\psi} {\mathcal D} \psi  \,+\, {\bar \varphi} {\mathcal
D} \varphi \right)} \;.
\end{equation}

This expression is invariant under two supersymmetry
transformations $s$, ${\bar s}$ that act on the fields as follows:
\begin{equation}
s \psi(x) \;=\;  \varphi(x) \;\;\;\; s{\bar \varphi}(x) \;=\;  {\bar \psi}(x) 
\end{equation}
\begin{equation}
{\bar s} {\bar \psi}(x) \;=\;  {\bar\varphi}(x) \;\;\;\; {\bar s}\varphi(x) \;=\; - \psi(x) 
\end{equation}
and all the other transformations equal zero.
It is evident that both $s$ and ${\bar s}$ are nilpotent 
($s$ and ${\bar s}$ are assumed to anticommute with $\psi$, ${\bar\psi}$).
The existence of these $BRST$-like symmetries implies, by the Parisi-Sourlas
mechanism~\cite{Parisi:1979ka}, that there are no physical field degrees of
freedom for the fields. Namely, the only remaining physical degrees
of freedom can only be zero-dimensional; i.e., particle-like.

Regarding the expression for the average over the random mass, we
point out to the fact that it is  possible, in principle, to
generalize the procedure to any even number of dimensions, since the Dirac
algebra may be also dealt with by the introduction of the proper
raising and lowering operators. 

\section*{Acknowledgments} 
The author thanks Prof. J. S\'anchez-Guill\'en for carefully reading the manuscript, and making useful comments.  He also acknowledges the financial support of CONICET (Argentina) and Fundaci{\'o}n Antorchas. 

\end{document}